\documentstyle{article}

\newcommand{\xtheta}{\Theta_0^0\hspace{-4ex}\raisebox{-1.8ex}{x-wave}}

\begin{document}
\begin{center}
{\Large\bf {Conformally coupled dark matter.} 
}\\[1cm]
{\bf
Mark Israelit}
\footnote{e-mail:
Marc.Israelit@uni-konstanz.de, and:  
israelit@physics.technion.ac.il, permanent e-address},
\\[1cm]
Department of Physics, University of Konstanz, PF 5560 M678,  
D-78434 Konstanz, Germany.
On leave from: Department of Physics, University of Haifa, Oranim,  
Tivon 36006, Israel.

\medskip

\end{center}

\begin{abstract}
\noindent
Dark matter is obtained from a scalar field coupled conformally to  
gravitation; the scalar being a relict of Dirac's gauge function.  
This conformally coupled dark matter includes a gas of very light  
($m\approx 2.25\times 10^{-34}\, eV$) neutral bosons having spin 0,  
as well as a time-dependent global scalar field, both pervading all  
of the cosmic space. The time-development of this dark matter in the  
expanding F-R-W universe is investigated, and an acceptable  
cosmological behaviour is obtained. 
\medskip

\noindent
{\bf Keywords}: General Relativity; Cosmology; Dark Matter.
\end{abstract}

\vspace{0.5cm}

\newpage

\section{Introduction.}
There is a great deal of dark matter in the universe, the dark  
matter having a much greater mean density then visible matter,  
(Trimble 1987), (Turner 1991), (Tremain 1992). It is believed (Tyson  
1992), (Priester, Hoell, \&  Blome 1995) that the density of dark  
matter exceeds that of luminous by a factor of about 10. At present  
there is a number of popular non-baryonic dark matter candidates:  
axions, neutrino, gravitino etc. etc. However, these candidates  
suffer from a number of shortcomings: their evolutionary behaviour  
in the expanding universe is not clearly known, their origin is not  
related to the geometry of the universe, they are inserted into the  
space-time framework from the outside.

Some years ago it was suggested (Israelit, \&  Rosen 1992) that at  
least a part of the dark matter in the universe consists of massive  
bosons of spin 1. These particles, named weylons were created by a  
local Weylian vector field. The idea of such a field was based on  
the Weyl geometry (Weyl 1919) as modified by Dirac (1973). It was  
also shown (Israelit, \&  Rosen 1994) that dark matter consisting of  
heavy weylons (with mass greater then 10 Mev), may be fitted into  
the standard cosmological model (Weinberg 1972), as well as into the  
singularity-free prematter cosmological model (Israelit, \&  Rosen  
1989). Later a global dark matter effect was obtained (Israelit, \&   
Rosen 1995) from a time-dependent scalar field, the scalar being  
identical with the gauge function introduced by Dirac (1973) in his  
modification of the Weyl geometry. The above mentioned dark matter  
forms are derived from the Weyl geometry. In this paper an  
additional geometrically based generator of cosmic dark matter is  
presented.

The purpose of the present work is to consider the possibility that  
a scalar field, coupled conformally to gravity, can generate cosmic  
dark matter. The behavior of this dark matter in the expanding  
universe will be investigated. It will also be shown that unifying  
the Einstein-Hilbert action with the (reduced) action of the  
Weyl-Dirac theory, one obtains the action of a scalar field coupled  
conformally to gravitation. In this procedure the coupled scalar  
field appears as a relict of the Dirac gauge function.

In the present work general relativistic (geometrical) units are  
used, so that $G=1, \quad c=1,$ the Hubble constant is given in  
$cm^{-1}$, and the energy density of matter in $cm^{-2}$. However,  
in order to discuss physical results we turn sometimes to  
conventional units, noting this fact in the text. The conversion  
rules may be found in (Synge 1966).


\section{The Weyl-Dirac theory and conformally coupling.}

A detailed description of the Weyl-Dirac theory and a discussion of  
its physical aspects may be found in Rosen's work (1982). A brief  
discussion of the theory in view of the dark matter problem is given  
in (Israelit, \&  Rosen 1992). In this section we consider the  
relation between the Weyl-Dirac theory and a scalar field coupled  
conformally to gravity. The notations of (Rosen 1982), and of  
(Israelit, \& Rosen 1992) will be used.

The field equations of the Weyl-Dirac theory may be derived from a  
variational principle (Dirac 1973), (Rosen 1982)

\begin {equation}\label{1}
\delta I_{D}=0,
\end{equation}
with the action
\begin {equation}\label{2}
I_{D}=\int L_{D} (-g)^{1/2} d^4x,
\end{equation}
and the Lagrangian density $L_{D}$ given by
\begin {equation}\label{3}
L_{D}=W_{\mu \nu} W^{\mu \nu}- \beta^2 R^{\sigma}_{\sigma}  
+k\beta_{\mu}\beta^{\mu}+(k-6)(2\beta w^{\mu} \beta_{\mu}
+\beta^2 w_{\mu}w^{\mu}) +2\Lambda \beta^4 \quad.
\end{equation}
In expression (\ref{3}), $w^{\mu}(x^{\lambda})$ is the connection  
vector of the Weylian geometry, the Weylian length curvature tensor  
$W_{\mu\nu}$ is defined as $W_{\mu\nu}=w_{\mu\, , \,\nu}-w_{\nu\, ,  
\, \mu}$ (a comma denoting partial differentiation),  
$\beta(x^{\lambda})$ is the Dirac gauge function, and
 $\beta_{\mu} \equiv \beta_{\, , \,\mu}$.
 Further, $\Lambda$ is the cosmological constant,  
$R^{\sigma}_{\sigma}$ is the Riemannian curvature scalar, and $k$ is  
an arbitrary parameter. In order to get
a geometrically based description of gravitation and  
electrodynamics, Dirac took $k=6$. In that case $w_{\mu}$ may be  
treated as the vector potential of the electromagnetic field, and  
$W_{\mu\nu}$ becomes the field tensor. By Dirac's choice one has  
from equation (\ref{3})
\begin {equation}\label{4}
L_{D}=W_{\mu\nu}W^{\mu\nu}-\beta^{\, 2} R^{\sigma}_{\sigma}+
6\beta^{\sigma}\beta_{\sigma}+2\Lambda \beta^{\, 4} \quad .
\end{equation}
For cosmological cosiderations we can neglect the Maxwell term in  
(\ref{4}), so that
\begin {equation}\label{5}
 L_{D}=-\beta^{\, 2} R^{\sigma}_{\sigma}+
6\beta^{\sigma}\beta_{\sigma}+2\Lambda \beta^{\, 4} \quad .
\end{equation}
Inserting (\ref{5}) into (\ref{2}) we obtain the action for a  
conformal space. Further, combining it with the Einstein-Hilbert  
action for gravitation and matter
\begin {equation}\label{6}
I_{G}=\int (R^{\sigma}_{\sigma}+L_{m}) (-g)^{1/2} d^4 x \quad,
\end{equation}
($L_{m}$ is the Lagrangian density of matter), we get
 \begin {equation}\label{7}
I=\int [R^{\sigma}_{\sigma}+L_{m}+8\pi(\beta^{\sigma}  
\beta_{\sigma}-\frac{1}{6}
\beta^2 R^{\sigma}_{\sigma}+\frac{1}{3} \Lambda \beta^4)](-g)^{1/2}  
d^4 x \quad.
\end{equation}
This is the general covariant action of a scalar field  
$\beta(x^{\lambda})$ coupled conformally to gravitation. In the  
following procedure $L_{m}$ will be considered as that of ordinary  
(luminous) matter, and it will be assumed that $L_{m}$ does not  
depend on $\beta$. On the other hand, dark matter will be obtained  
from the conformally coupled scalar field $\beta(x)$. We also shall  
consider models without the cosmological constant so that in the  
action (\ref{7}) we shall set  $\Lambda=0$.

\section{The field equations. }
Varying in(\ref{7}) the metric tensor $g_{\mu\nu}$ we obtain the  
Einstein equation
 \begin {equation}\label{8}
R_{\mu\nu}-1/2~ g_{\mu\nu} R_{\sigma}^{\sigma}=-8\pi(T_{\mu\nu}+
\Theta_{\mu\nu}) \quad,
\end{equation}
with the Ricci tensor $R_{\mu\nu}$, and with the energy-momentum  
density tensor
$T_{\mu\nu}$ of ordinary (luminous) matter
\begin {equation}\label{9}
8\pi \sqrt{-g}~ T_{\mu\nu}=\frac{\delta(\sqrt{-g}L_{m}}{\delta  
g^{\mu\nu}} \quad.
\end{equation}
The dark matter is represented in (\ref{8}) by the energy-momentum  
density tensor of the $\beta$-field
\begin {equation}\label{10}
6\Theta_{\mu\nu}=4\beta_{\mu}\beta_{\nu}-g_{\mu\nu}\beta_{\sigma}
\beta^{\sigma}-2\beta \beta_{\mu\, ;\, \nu}+2g_{\mu\nu}\beta
\beta^{\sigma}_{\, ;\, \sigma}-\beta^2(R_{\mu\nu}-1/2~g_{\mu\nu}
R^{\sigma}_{\sigma})\quad.
\end{equation}
Further, variation with respect to the field $\beta$ yields the  
following field equation:
\begin {equation}\label{11}
\beta^{\sigma}_{\,;\,\sigma}+(1/6)\beta R^{\sigma}_{\sigma}=0.
\end{equation}
Making use of (\ref{11}), one obtains from (\ref{10})
\begin {equation}\label{12}
\Theta^{\sigma}_{\sigma}=0\quad,
\end{equation}
so that (\ref{8}) gives
\begin {equation}\label{13}
R^{\sigma}_{\sigma}=8\pi T^{\sigma}_{\sigma} \quad.
\end{equation}
Taking into account (\ref{13}), one can rewrite the field equation  
(\ref{11}) as
\begin {equation}\label{14}
\beta^{\sigma}_{\,;\,\sigma}=-\frac{4\pi}{3}~ \beta ~  
T^{\sigma}_{\sigma} \quad,
\end{equation}
so that the behavior of the conformally coupled field depends on  
the amount and state of ordinary matter. Finally, making use of the  
contracted Bianchi identity and of (\ref{11}), one obtains from  
(\ref{10})
\begin {equation}\label{15}
\Theta^{\nu}_{\mu\, ;\,\nu}=0 \quad,
\end{equation}
and from (\ref{8}) one has
\begin {equation}\label{16}
T^{\,\nu}_{\mu\, ;\,\nu}=0 \quad,
\end{equation}
so that one obtains separated energy-momentum relations, the first,  
(\ref{15}) for dark matter, and the second, (\ref{16}) for ordinary  
matter.

Let us consider a homogeneous and isotropic spatially closed   
universe described by the F-R-W line element
\begin {equation}\label{17}
ds^2=dt^2-R^2\left(\frac{dr^2}{1-r^2}+r^2d\vartheta^2+r^2\sin^2\vartheta  
~d\varphi^2
\right) \quad,
\end{equation}
with $R(t)$ being the radius of the universe. We assume that the  
universe is filled with ordinary cosmic matter, having an energy  
density $\rho(t)$, and a pressure $P(t)$, as well as with dark  
matter, created by the scalar function $\beta$. For the metric given  
by (\ref{17}) we can write down the Einstein equation (\ref{8})  
with $\mu=0; \quad \nu=0;$
\begin {equation}\label{18}
\frac{\dot{R}^2}{R^2}+\frac{1}{R^2}=\frac{8\pi}{3}(\rho+\Theta^0_0)  
\quad,
\end{equation}
and from eq. (\ref{13}) we obtain
\begin {equation}\label{19}
\frac{\ddot{R}}{R}+\frac{\dot{R}^2}{R^2}+\frac{1}{R^2}=\frac{4\pi}{3}(\rho-
3P) \quad.
\end{equation}
In addition we have from the energy relation (\ref{16})
\begin {equation}\label{20}
\dot{\rho}+\frac{3 \dot{R}} {R}(\rho+P)=0.
\end{equation}
The equations obtained in this section enable us to investigate the  
scalar field $\beta$ coupled conformally to a F-R-W universe filled  
with ordinary matter. The generation of dark matter by this  
$\beta$-field may be considered either by means of a local approach,  
or by a global one.

\section{A local approach. Scalar bosons.}
From the global standpoint, in a homogeneous and isotropic universe  
$\beta$ depends only on the cosmic time $t$. However, one can  
assume that there exist small regions of inhomogeneity, in which  
$\beta$ may depend also on spatial coordinates. Introducing local  
Lorentzian coordinates ($c\tau,x,y,z$), one can rewrite (\ref{14})  
as
\begin {equation}\label{21}
\Box \beta+\frac{4\pi}{3}(\rho-3P)\beta=0.
\end{equation}
Further in small space-time regions one can neglect the dependence  
of $\rho$ and $P$ on the cosmic time $t$, so that
\begin {equation}\label{22}
\frac{4\pi}{3}(\rho-3P)\equiv\kappa^2 \simeq~ const.\quad,
\end{equation}
and (\ref{21}) may be rewritten in the form of a Klein-Gordon equation
\begin {equation}\label{23}
\Box \beta+\kappa^2\beta=0.
\end{equation}
From the point of view of quantum mechanics, eq. (\ref{23})  
describes spinless particles, scalar bosons, with a mass $m$ and  
Compton wavelength $\lambda_c$. In conventional units one has
\begin {equation}\label{24}
\kappa=\frac{mc}{\hbar}=\frac{2\pi}{\lambda_c}~,
\end{equation}
and by (\ref{22}), and (\ref{24}) one obtains for the mass in  
conventional units
\begin {equation}\label{25}
m=\frac{\hbar}{c^3}\left[\frac{4\pi}{3}G(\rho-3P)\right]^{1/2} \quad,
\end{equation}
with $G$ being Newton's gravitational constant. It will be recalled  
that $\rho$, and $P$ are the energy density and pressure of  
ordinary (luminous) cosmic matter.

In order to estimate the boson mass at present we will make use of  
the Hubble constant. Following (Peebles 1993) let us write for its  
present value ~
$H_{_N}=H_{0}h$, where~
$H_{0}=100~km~s^{-1}~Mpc^{-1}\doteq(9.78\times10^9~y)^{-1}\doteq
1.08\times10^{-28}~cm^{-1}$~ ({cf. (Weinberg 1972)), and with
 $0.5\leq~h\leq~0.8$ ~ (cf.(Priester, Hoell,\&  Blome 1995)) .   
Adopting the value
$h\approx0.5$ we get
\begin {equation}\label{26}
H_{_N}=0.5\times10^{-28} ~ cm^{-1},
\end{equation}
so that for the critic density we obtain
\begin {equation}\label{27}
\rho_{_C}=\left(\frac{3}{8\pi}\right)~H_{_N}^2=0.3\times 10^{-57}  
cm^{-2}.
\end{equation}
To discuss a closed universe~ one can take for the total (~luminous  
and dark~) matter density at present~ ~  
$\rho_{total~N}\doteq0.35\times10^{-57}~cm^{-2}>\rho_{c}$. Further  
one can assume that the luminous matter makes up 0.1 of the total  
one, so that the following values must be substituted  into  
(\ref{25}):
\begin {equation}\label{28}
P_{_N}=0 ~, \quad \rho_{_{N}}=0.35\times10^{-58}~cm^{-2}~ .
\end{equation}
Expressing $\rho_{_N}$ in conventional units one obtains the  
present value
\begin {equation}\label{29}
m_{_{N}}\approx 4\times10^{-67}=2.25\times10^{-34}~ eV~~.
\end{equation}
According to(\ref{25}) in the radiation period the boson mass  
vanishes, whereas during the dust-dominated period it is given by
\begin {equation}\label{30}
m=m_{_{N}} \left(\frac{R_{_N}}{R}\right)^{1/2}~~,
\end{equation}
with $R_{_N}$ being the present value of the cosmic scale factor.

Let us consider the energy-momentum density of the dark matter. For  
small regions, in local Lorentzian coordinates, one obtains from  
(\ref{10}) the following non-vanishing components:
\begin {equation}\label{31}
\Theta^0_0=\frac{1}{2  
c^2}\beta_{\tau}^2+\frac{1}{6}(\beta_{x}^2+\beta_{y}^2
+\beta_{z}^2)-\frac{1}{3}\beta (\beta_{xx}+\beta_{yy}+\beta_{zz})+
\frac{1}{2} \beta^2\left(\frac{\dot{R}^2}{R^2}+\frac{1}{R^2}\right)  
\quad,
\end{equation}
\begin {equation}\label{32}
\Theta^1_0=\frac{2}{3 c}\beta_{\tau}\beta_{x}+\frac{1}{3  
c}\beta\beta_{\tau x}
\end{equation}\quad,
\begin {eqnarray}\label{33}
\Theta_1^1=-\frac{1}{2}  
\beta_x^2+\frac{1}{6}\left(\beta_{y}^2+\beta_{z}^2
-\frac{1}{c^2}\beta_{\tau}^2\right)-\frac{1}{3}\beta  
\left(\beta_{yy}+\beta_{zz}-\frac{1}{c^2}\beta_{\tau \tau}\right)
\\
\nonumber
+\frac{1}{6} \beta^2\left(\frac{2~\ddot{R}}{R}  
+\frac{\dot{R}^2}{R^2}+\frac{1}{R^2}\right) \quad.
\end{eqnarray}
and similar expressions for $\Theta_0^2 ~, \Theta_2^2~,  
\Theta_0^3~, \Theta_3^3$.
In order to calculate the energy density and pressure of dark  
matter, let us consider a plane wave traveling in the $x$-direction
\begin{equation}\label{34}
\beta=B\cos (\omega \tau-kx) ~; \quad (B, \omega,k=const.)
\end{equation}
Making use of (\ref{34}) we get from (\ref{23})
\begin{equation}\label{35}
\frac{\omega^2}{c^2}=k^2+\kappa^2 .
\end{equation}
In the quantum-mechanical representation one writes
\begin{equation}\label{36}
\varepsilon=\hbar \omega~, \qquad p=\hbar k~.
\end{equation}
where $\varepsilon$,  and $p$ are the energy and momentum of the  
particle, so that with (\ref{24}) relation (\ref {35}) may be  
written as

\begin{equation}\label{37}
\varepsilon^2= c^2 p^2 +m^2c^4 ~~.
\end{equation}
For the plane wave given by (\ref{34}) we obtain from  (\ref{31})
\begin{equation}\label{38}
\xtheta=(1/2)~\beta^2  
\left[\frac{\omega^2}{c^2}+k^2+\left(\frac{\dot{R}^2}
{R^2}+\frac{1}{R^2} \right)\right] ~~.
\end{equation}
To describe an ensemble of particles having spatial isotropy, we  
can take a combination of six plane waves: the wave (\ref{34})  
moving in the positive direction allong the $x$-axis, one moving in  
the opposite direction, two waves moving along the $y$-axis in both  
directions, and two moving along the $z$-axis. If we take the time  
average, we obtain for the combined system of waves
\begin{equation}\label{39}
\overline\Theta^0_0=(3/2)~B^2  
\left(\frac{\omega^2}{c^2}+k^2+\frac{\dot{R}^2}{R^2}+\frac{1}{R^2}\right)~~.
\end{equation}
Making use of (\ref{32}), we obtain for the same system of six  
plane waves
\begin{equation}\label{40}
\overline\Theta^1_0=\overline\Theta^2_0=\overline\Theta^3_0=0~~.
\end{equation}
Finally, making use of (\ref{18}), (\ref{22}), and (\ref{35}), we  
get from equation (\ref{33})
\begin{equation}\label{41}
\overline\Theta^1_1=\overline\Theta^2_2=\overline\Theta^3_3=
-\frac{1}{3}\overline\Theta^0_0~~.
\end{equation}
The expression (\ref{39}) is actually the energy density of our  
dark matter $\rho_{_{D}}$. One can consider it as consisting of two  
parts, the first being the contribution $\rho_{_{B}} $ of scalar  
bosons, the second, $\rho_{_{G}}$,
representing a global effect.
\begin{equation}\label{42}
\rho_{_{B}}=(3/2)~B^2\left(\frac{\omega^2}{c^2}+k^2\right)~, \quad
\rho_{_{G}}=(3/2)~B^2\left(\frac{\dot{R}^2}{R^2}+\frac{1}{R^2}\right)~,
\quad \rho_{_{D}}\equiv\overline\Theta^0_0=\rho_{_{B}}+\rho_{_{G}}~~.
\end{equation}

Up to here a monoenergetic ensemble of bosons is discussed.  
According to (\ref{25}) the boson has zero mass in the radiation  
era, whereas in the dust period it is extremely light (cf.  
(\ref{29}), and (\ref{30})).These particles are discribed by the  
Bose-Einstein statistics, and there is no interaction with ordinary  
matter, so that the total number of bosons in the universe is  
conserved. In order to get a realistic scenario, one can turn to a  
gas of
scalar bosons in thermal equilibrium, and fit it into the  
cosmological model. This procedure was already carried out in  
(Israelit \& Rosen 1992, 1994) for a gas of weylons.

\section{A global approach.}
Above a F-R-W universe containing small regions of local  
inhomogeneities was considered. In the present section we consider a  
completely homogeneous and isotropic universe, with $\beta$  
depending only on the cosmic time $t$. With the metric given by  
(\ref{17}), we can rewrite eq. (\ref{14}) as
\begin{equation}\label{43}
\left(\dot{\beta} R^3 \right)_{ \, , \,t}=- \frac{4\pi}{3}(\rho-3P)  
R^3\beta \quad.
\end{equation}
Further, making use of (\ref{17}), and (\ref{43}), we obtain from  
(\ref{10}) the following non-vanishing components of the  
energy-momentum density tensor of the global scalar field
\begin{equation}\label{44}
\Theta^0_0=\frac{1}{2} ~\dot{\beta}^2  
+\frac{\dot{R}\dot{\beta}\beta}{R}+\frac{1}{2}~\beta^2
\left(\frac{\dot{R^2}}{R^2}+\frac{1}{R^2}\right)~~,
\end{equation}
\begin{equation}\label{45}
\Theta^1_1=\Theta^2_2=\Theta^3_3= -\frac{1}{3}\Theta^0_0 \quad.
\end{equation}
One can consider $\beta$ as a function of the cosmic scale factor  
$R$, so that
\begin{equation}\label{46}
\dot{\beta}=\dot{R} \beta _{\, ,\, R} \quad \rm{with} \quad \beta  
_{\, ,\, R}\equiv
\frac{d\beta}{dR} \quad,
\end{equation}
and (\ref{44}) may be rewritten as

\begin{equation}\label{47}
\rho_{_{D}}\equiv\Theta^0_0=\frac{\dot{R}^2}{2 R^2}  
\left(\beta+R\beta_{\,,\,R}\right)^2+\frac{\beta^2}{2R^2} \quad.
\end{equation}
It is remarkable that in this approach, like in the previous one  
(cf.(\ref{39})), the energy density $\rho _{_{D}}$ of the scalar  
field is positive, so that one obtains a dark matter effect.

Making use of (\ref{46}) we can rewrite eq. (\ref{43}) as follows
\begin{equation}\label{48}
R^3 \dot{R}^2 \beta_{\,,R\,,\,R}+R^2(R\ddot{R}+3 \dot{R}^2 )  
\beta_{\, , \, R}+
\frac{4\pi}{3} R^3(\rho-3P) \beta =0 \quad.
\end{equation}
For a moment, let us go back to the relations (\ref{12}), and  
(\ref{15}). Making use of (\ref{17}) we obtain for the energy  
density of dark matter
\begin{equation}\label{49}
\frac{8\pi}{3} \rho_{_D}\equiv\frac{8\pi}{3}  
\Theta_0^0=\frac{D}{R^4}~~; \qquad
(D=\rm const.).
\end{equation}

Let us consider the dust-dominated period. Here $P \ll \rho$, and  
the energy relation (\ref{20}) gives
\begin{equation}\label{50}
\frac{8\pi}{3} \rho=\frac{M}{R^3} ~~; \qquad (M=\rm const.).
\end{equation}
Substituting (\ref{18}), and (\ref{19}) into (\ref{48}), and taking into 
account (\ref{49}), (\ref{50}), we obtain
\begin{equation}\label{51}
\left[MR^2+D R -R^3\right] \beta_{ , R ,  R}+
\left[\frac{5}{2}M R+2D-3R^2 \right]\beta_{, R}=-\frac{1}{2}M  
\beta~\quad,
\end{equation}
and turning to the dimensionless variable
\begin{equation}\label{52}
z=\frac{R}{R_{_N}} \quad,
\end{equation}
with $R_N$ being the present value of the cosmic scale factor, we get
\begin{equation}\label{53}
\left[M R_{_N} z^2+D z-R^2_{_N}  
z^3\right]\beta_{,z,z}+\left[\frac{5}{2}M R_{_N} z
+2D-3R^2_{_N} z^2\right) \beta_{,z}+\frac{1}{2}M R_{_N} \beta =0 ~~~.
\end{equation}
The constants $M$, $D$, and $R_{_N}$ apearing in (\ref{53}) may be  
expressed in terms of two observable cosmological quantities, the  
present value of the Hubble "constant"
 $H_{_N}$, and that of the energy density of ordinary (luminous)  
matter $\rho_{_N}$. It is convenient to introduce a parameter  
$\eta$~, given by
\begin{equation}\label{54}
R_{_N}=\frac{\eta}{H_{_N}}~~.
\end{equation}
A second helpful parameter is the ratio
\begin{equation}\label{55}
\chi=\frac{\rho_{_{total}}}{\rho}\equiv 1+\frac{\rho_{_D}}{\rho} \quad ,
\end{equation}
so that for the present state of the universe we have
\begin{equation}\label{56}
\rho_{_{total, N}}=\rho_{_N}+\rho_{_{D,N}}=\chi_{_N} \rho_{_N}~~.
\end{equation}

On the other hand, eq. (\ref{18}) at present may be written as
\begin{equation}\label{57}
H^{2}_{_N}+\frac{1}{R^{2}_{_N}}=\frac{8\pi}{3}(\rho_{_N}+\rho_{_{D,N}})~~.
\end{equation}
With (\ref{27}), and (\ref{54}) this takes on the form

\begin{equation}\label{58}
\rho_{_{total}}=\rho_{_N}+\rho_{_{D,N}}=\rho_{_C}\left(1+\frac{1}{\eta^2}\right)\quad
\end{equation}
Finally, making in (\ref{58}) use of (\ref{27}), (\ref{49}),  
(\ref{50}), and (\ref{54}), we obtain
\begin{equation}\label{59}
(1+\eta^2) R^2_{_N}=MR_{_N}+D~~.
\end{equation}
In a similar way we obtain from (\ref{56})
\begin{equation}\label{60}
\chi_{_N} M R_{_N}=MR_{_N}+D ~~.
\end{equation}
From (\ref{59}), and (\ref{60}) one obtains
\begin{equation}\label{61}
M=\frac{1+\eta^2}{\chi_{_N}}R_{_N}~~,
\end{equation}
and
\begin{equation}\label{62}
D=\frac{\chi_{_N}-1}{\chi_{_N}}(1+\eta^2)R^2_{_N}~~~.
\end{equation}
Inserting (\ref{61}), and (\ref{62}) into eq. (\ref{53}) we get:
\begin{eqnarray}\label{63}
&\left[z^3-\frac{1+\eta^2}{\chi_{_N}}~z^2-\frac{\chi_{_N}-1}{\chi_{_N}}~  
  (1+\eta^2)~z\right]~\beta_{,z,z}&
\\
\nonumber
+&\left[3z^2-\frac{5(1+\eta^2)}{2\chi_{_N}}~z-\frac{2(\chi_{_N}-1)}{\chi_{_N}}~
(1+\eta^2)\right]~\beta_{,z}&
-\frac{1+\eta^2}{2\chi_{_N}}~\beta=0~~.
\end{eqnarray}
For given values of parameters $\eta$, and $\chi_{_N}$ one can  
obtain the dark matter function~~ $\beta(z)$ from equation  
(\ref{63}). It is believed (Tyson 1992),
(Blome, Priester,\&  Hoell 1995) that at present the total density  
of matter exceeds that of luminous by a factor of $\sim  10$. This  
justifies taking
\begin{equation}\label{64}
\chi_{_N}=10 ~~.
\end{equation}
From (\ref{54}), and (\ref{58}) we see that large values of $\eta$  
describe a nearly flat universe. Below will be considered models  
with $\eta=1,~4,~ 20$, combined with $\chi_{_N}=10$.

Let us take the Einstein equation (\ref{18}). Making use of  
(\ref{49}), and (\ref{50}), we can rewrite it as
\begin{equation}\label{65}
\frac{\dot{R}^2}{R^2}+\frac{1}{R^2}=\frac{M}{R^3}+\frac{D}{R^4} ~~~.
\end{equation}
A closed universe achieves its maximum radius at $\dot{R}=0$, so  
that from (\ref{65}) we can calculate $R_{max}$.  Further,  
substituting (\ref{49}) into(\ref{47}), we obtain
\begin{equation}\label{66}
\frac{\dot{R}^2}{R^2}(\beta+R\beta_{,R})^2+\frac{\beta^2}{R^2}=
\frac{3}{4\pi}~\frac{D}{R^4}~~.
\end{equation}
Making use of the values of $R_{max}$ obtained from (\ref{65}), one  
can calculate $\beta_{max}\equiv\beta(R_{max})$, the value at the  
state of maximum expansion. Finally, taking into account (\ref{49}),  
and (\ref{50}), as well as (\ref{61}), and (\ref{62}) we have for  
the ratio (\ref{55}) during the dust-dominated period
\begin{equation}\label{67}
\chi (R)=1+\frac{D}{MR}=1+\frac{(\chi_{_N}-1)R_{_N}}{R}~~,
\end{equation}
so that
\begin{equation}\label{68}
\chi_{max}=1+\frac {(\chi_{_N}-1)R_{_N}}{R_{max}}~~~.
\end{equation}
Taking  the values of  $R_{max}$  obtained from eq. (\ref{65}), we  
can from (\ref{68}) calculate the values of $\chi_{max}$. For a  
moment let us go back to (\ref{51}). At the state of maximum  
expansion the first bracket in that equation vanishes  
(cf.(\ref{65})). Thus, making use of (\ref{61}), and (\ref{62}), and  
of the above mentioned values of $\beta_{max}$, one can calculate  
the value of the derivative $\left(\beta_{,R}\right)_{max}$, for the  
moment when $R=R_{max}$. By the procedure described above, we get  
from equations (\ref{51}), (\ref{65}),  (\ref{66}), (\ref{68}), for  
chosen values of $\chi_{_N}$ and $\eta$, the results listed below:    

\begin{eqnarray}\label{69}
&\chi_{_N}=10;~\eta=1;~R_{max}=1.4R_{_N};~\beta_{max}=0.45;~  
\chi_{max}=7.4;
\left(\beta_{,R}\right)_{max}=2.3\times10^{-2}H_{_N}&
\\
\nonumber
&\chi_{_N}=10;~\eta=4;~R_{max}=4.8R_{_N};~\beta_{max}=0.39;~  
\chi_{max}=2.9;
\left(\beta_{,R}\right)_{max}=4.3\times10^{-4}H_{_N}&
\\
\nonumber
&\chi_{_N}=10;~\eta=20;~R_{max}=48R_{_N};~\beta_{max}=0.19;~  
\chi_{max}=1.2;
\left(\beta_{,R}\right)_{max}=1.2\times10^{-4}H_{_N} &
\end{eqnarray}
From (\ref{58}) one can readily see that the first line in  
(\ref{69}) belongs to a well closed universe, whereas the third line  
describes an almost flat model. It must be noted that we have made  
use of two equations, (\ref{51}), and (\ref{66}), both for the same  
function  $\beta(R)$. However differentiating
(\ref{66}) we can verify that these two equations are coexistent.
Now, if one wants, he can obtain $\beta(R)$ from the linear  
differential equation (\ref{51}) (or alternatively from (\ref{63}))  
for a chosen pair of parameters $\chi_{_N}$, $\eta$.  In this  
procedure the above given values of
$\beta_{max}\equiv\beta(R_{max})$, and  
$\left(\beta_{,R}\right)_{max}$, serve as boundary conditions.

Let us consider equation (\ref{65}). Integrating it one obtains
\begin{equation}\label{70}
t+C=-\left(D+MR-R^2\right)^{1/2}-\frac{M}{2}\sin^{-1}\frac{M-2R}{\sqrt{M^2+4D}}~~,
\end{equation}
with $C$ being a constant. Inserting  (\ref{61}), and (\ref{62}),  
turning to $z$ according to (\ref{52}), and making use of  
(\ref{54}), one can rewrite
(\ref{70}) as follows
\begin{equation}\label{71}
t+C=-\frac{\eta}{H_{_N}}
\left\{\left[\frac{1+\eta^2}{\chi_{_N}}(\chi_{_N}-1+z)
-z^2\right]^{1/2}
+\frac{1+\eta^2}{2\chi_{_N}}\sin^{-1}
\frac{1+\eta^2-2\chi_{_N}z}
{(1+\eta^2)\sqrt{1+\frac{4(\chi_{_N}-1)\chi_{_N}}{1+\eta^2}}}
\right\}~.
\end{equation}

With the help of (\ref{71}) one can estimate the age $\tau_{_N}$ of  
the universe. For this purpose one neglects the duration of the  
radiation, and preradiation periods (cf. (Weinberg 1973), (Israelit,  
\& Rosen 1989)).  One also must take into account the fact that the  
radius of the universe at the beginning of the dust-dominated  
period $R_0$ is much smaller then its present value $R_{_N}$,  ($R_0  
<  0.01R_{_N}$). Further, making use of the values of $R_{max}$  
quoted in  (\ref{69}), one can also estimate the moment
$\tau_{max}$, when the universe reaches its maximum radius.  
Straightforward calculations give the following results.
\begin{eqnarray}\label{72}
&\chi_{_N}=10;~~\eta=1;~~ \Rightarrow~~ \tau_{_N}=0.42H^{-1}_{_N};~ ~  
\tau_{max}=1.13H^{-1}_{_N}&~~.
\\
\nonumber
&\chi_{_N}=10;~~\eta=4;~~ \Rightarrow~~ \tau_{_N}=0.50H^{-1}_{_N};~ ~  
\tau_{max}=14.2H^{-1}_{_N}&~~.
\\
\nonumber
&\chi_{_N}=10;~~\eta=20; ~~\Rightarrow~~ \tau_{_N}=0.51H^{-1}_{_N};~ ~  
\tau_{max}=144H^{-1}_{_N}&~~.
\end{eqnarray}
Let us take for the present value of the Hubble constant (cf. (\ref{26}))
\begin{equation}\label{73}
H_{_N}=50~km~s^{-1}~Mpc^{-1}\doteq(2\times10^{10}~y)^{-1}\doteq
0.5\times10^{-28}~cm^{-1}~.
\end{equation}
Then we obtain from (\ref{72}) for the models considered above
\begin{equation}\label{74}
0.84 \times10^{28} \rm{cm}\leq \tau_{_N}\leq 1.02 \times 10^{28} \rm{cm}
\end{equation}
and turning to time units we can rewrite this as
\begin{equation}\label{75}
8.9\times10^{9}\rm{y}\leq \tau_{_N}\leq 10.2\times10^{9}\rm{y}~~.
\end{equation}

\section{Discussion.}
In the present work dark matter is obtained from a conformally  
coupled scalar field. It is shown that this field may be treated as  
a source of scalar bosons and that it can  also cause a global dark  
matter effect. In the process of creating dark matter the role of  
the conformally coupled scalar field is similar to that of the Dirac  
gauge function. However, whereas it was possible to choose the  
latter almost arbitrarily (cf. (Israelit, \& Rosen 1995)), the  
behavior of the conformally coupled field depends on the density and  
pressure of ordinary (luminous) matter in the universe  (cf.  
equations (\ref{14}), (\ref{21}), (\ref{43})).

Two aproaches are considered, a local, and a global one. In the  
first case we assumed that the globally homogeneous and isotropic  
universe contains small regions of local inhomogeneities, wherer the  
scalar function $\beta$ may depend on spatial coordinates. This  
dependence leads to a Klein-Gordon equation for $\beta$ (eq.  
(\ref{23})), so that from the quatum mechanical standpoint one  
obtains an ensemble of neutral and spinless bosons. During the  
radiation period these particles have zero mass, and in the  
dust-dominated universe they are extremely light, so that these  
particles constitute dark matter pervading all of the cosmic space.

In the global approach (section 5.) the universe is completely  
homogeneous and isotropic, and the scalar function $\beta(t)$  
generates a global dark matter effect. In order to specify  
cosmological models, two parameters are introduced, the first,  
$\eta$ characterizing the degree of flatness of the universe, and a  
second, $\chi$ being the ratio of the total matter density to that  
of ordinary (luminous) matter. It is assumed that at present  
$\chi_{_N}=10$, and for various values of  $\eta$ the present age of  
the universe, its maximum radius, as well as $\tau_{max}$, the  
moment when the expanding universe will achieve its maximum radius  
are calculated. From (\ref{69}), (\ref{72}-{75}) one can conclude  
that the model with   $\chi_{_N}=10$ and $\eta=20$ (an almost flat  
universe) agrees with nowadays cosmological observations (Bolte, \&   
Hogan 1995), (Chaboyer, Demarque, Kernan, \& Krauss 1996). For this  
model we have:
$R_{max}=48R_{_N};~~\tau_{_N}=10.2\times10^9 \rm{y}; ~~\tau_{max}    
        =282\tau_{_N}$. In the expanding universe, during the  
dust-dominated period, the ratio $\eta$ (cf. (\ref{67})) decreases  
by a factor of about 1000, hence this dark matter form will not play  
an essential role in the late stages of the expansion phase, but it  
was important in the early stages of the dust universe. So it might  
have affected growing processes of galactic formations at the  
beginning of the dust era (Israelit, Rose, \& Dehnen 1994,a,b).

The conformally coupled dark matter considered in the present paper  
is to be regarded as a dark matter form connected closely to the  
Weyl geometry.  Assuming that the space-time of the universe has  
Weylian properties (in addition to the well known Riemannian  
properties) we can derive several geometrically based dark matter  
forms (cf. (Israelit 1995)): a) the dark matter considered in this  
work,  b) a gas of heavy weylons (Israelit, \& Rosen 1992, 1994),   
c) a global dark matter effect from the Dirac gauge function  
(Israelit, \& Rosen 1995). All these forms may be fitted into  
cosmological models, giving agreement with observation.


\section*{Acknowledgments}
{ The author takes this opportunity to express his cordial thanks  
to Professor HEINZ DEHNEN for interesting discussions.}

{\bf REFERENCES.}

Blome,~H.J., Priester,~W.,\&  Hoell,~J.: 1995, {\it Currents in  
High Energy Astrophysics}, p.-p. 301-312,  Shapiro et al. (eds),  
(Kluver Academic Publisher).

Bolte,~M, \& Hogan,~C.,J. : 1995, {\it Nature}, {\bf 376}, 399.

Chaboyer,~B., Demarque,~P., Kernan,~P.J.,  \& Krauss,~L.,M. : 1996,  
{\it Science}, {\bf 271}, 957 .

Dirac,~P.,~M.,~A., : 1973,  {\it Proc. R. Soc. London} {\bf A 333},  
403 .

Israelit,~M.,  1995, {\it Proceedings of the Third Alexander  
Friedman International Seminar on Gravitation and Cosmology},   
Gnedin et al. (eds)
(Friedmann Laboratory Publishing, St. Petersburg)~ p.-p. 126-143 .

Israelit,~M., Rose,~B., \& Dehnen,~H., : 1994 a , {\it Astrophys.  
Space Sci},
{\bf 219}, 171 .

Israelit,~M., Rose,~B., \& Dehnen,~H., : 1994 b , {\it Astrophys.  
Space Sci},
{\bf 220}, 39 .

Israelit,~M., \&  Rosen,~N., : 1989, {\it Astrophys. J.}, {\bf  
342}, 627 .

Israelit,~M., \&  Rosen,~N., : 1992,  {\it Found. Phys.}, {\bf 22}, 555 .

Israelit,~M., \&  Rosen,~N., : 1994,  {\it Found. Phys.}, {\bf 24}, 901 .

Israelit,~M., \&  Rosen,~N., : 1995,  {\it Found. Phys.}, {\bf 25}, 763 .

Peebles,~P.,~J.,~E., 1993, {\it Principles of Physical Cosmology},  
(Princeton University Press, Princeton, New Jersey).

Priester,~W.,  Hoell,~J., \&  Blome,~H.,~J., : 1995, {\it Comments  
Astrophys.},
{\bf 17}, No.6, 327 .

Rosen,~N., : 1982, {\it Found. Phys.}, {\bf 12}, 213 .

Synge,~J.,~L., : 1966, {\it Relativity, the General Teory},  
(North-Holland Publ. Comp., Amsterdam) CH.IV-5, and Appendix B .

Tremain,~S.,: 1992, {\it Phys. Today} , {\bf 45}, No.2, 28 .

Trimble,~ V., : 1987, {\it Ann. Rev. Astron. Astrophys.} , {\bf  
25}, 425 .

Turner,~M.,~S., : 1991, {\it Physica Scripta},  {\bf T 36}, 167 .

Tyson,~A., : 1992, {\it Phys. Today}, {\bf 45}, No.6, 24 .

Weiberg,~S., : 1972, {\it Gravitation and Cosmology}, (Wiley, New York).

Weyl,~H., : 1919, {\it Ann.Phys. (Leipzig)},  {\bf 59} , 101 .

\end{document}